\documentclass[12pt]{article} 
\usepackage{epsfig} 
 
\newcommand{\be}{\begin{equation}} 
\newcommand{\ee}{\end{equation}} 
\newcommand{\bea}{\begin{eqnarray}} 
\newcommand{\eea}{\end{eqnarray}}

\begin{document} 
\bibliographystyle{unsrt} 
 
\title{Crackling Noise, Power Spectra and Disorder Induced Critical Scaling}

\author{A. Travesset, R.A. White, and K.A. Dahmen \\
Loomis Laboratory, University of Illinois at Urbana\\
61801, Urbana, IL, USA\\}
 
\date{}

\maketitle 
 
\begin{abstract} 
Crackling noise is observed in many 
disordered non-equilibrium systems in response
to slowly changing external conditions.
Examples range from Barkhausen noise in magnets to acoustic emission
in martensites to earthquakes.
Using the non-equilibrium
random field Ising model, we derive universal scaling predictions
for the dependence of the associated power spectra on the disorder and field
sweep rate, near an underlying disorder-induced non-equilibrium critical 
point. Our theory applies to certain systems
in which the crackling noise results from 
avalanche-like response to a (slowly) increasing external driving force,
and is characterized by a broad power
law scaling regime of the power spectra. We compute the critical exponents 
and discuss the relevance of the results to experiments.
\end{abstract} 
 
\vfill\newpage 
 
\pagebreak 
 
\section{Introduction}\label{SECT__Int}

Many non-equilibrium physical systems ranging from disordered ferromagnets
to superconducting vortices \cite{FWNL:95}, to 
martensitic shape-memory alloys \cite{VIV1:94,VIV2:95,VIV3:95,VIV4:98}, 
to earthquakes, exhibit crackling 
noise\cite{SDM:01}
in response to smoothly varying external conditions (driving force).
While it is not at all obvious that these systems should show similar 
properties, it is observed
that these and other examples in nature exhibit power law scaling of the 
noise statistics
over many decades. The exponents 
observed in these systems fall into distinct 
universality classes:
Barkhausen noise observed in disordered ferromagnets exhibits power 
law scaling in the power spectrum (PS) 
\be\label{zero_equation}
P(\omega) \sim \omega^{-1/\sigma\nu z} \ ,
\ee
for large frequencies $\omega$, with a universal exponent
$1/\sigma\nu z$ that does not depend on the microscopic details,
but only on a few basic properties such as symmetries, dimensions,
range of interaction, etc. Since universal power law characteristics are
often associated with systems at or near an underlying critical point, models with critical 
points have been suggested to understand the origin of the observed universality.

Barkhausen noise, the characteristic crackling noise associated
with the motion of magnetic domain walls in a ferromagnet as the  
external magnetic field is slowly varied, has enjoyed a significant amount of
experimental and theoretical attention in the past decade because it is a 
particularly simple and experimentally readily accessible example of systems
with crackling noise. Models for Barkhausen noise in soft ferromagnets
usually model the motion of one (or few) domain walls,
while other models, presumably for Barkhausen noise in hard (strongly disordered) magnets 
model the collective behavior of many interacting domain walls.
While the former is marked by long range ferromagnetic interactions which cause
the system to ``self tune'' to the domain wall 
depinning transition \cite{Zap:97} the latter is not self tuned
and requires a different mechanism to explain scaling. This mechanism comes in the form of 
a disorder induced critical point found in the zero temperature, 
random field Ising model (RFIM) \cite{BKD:98,JRob:92} 
studied far from thermal equilibrium \cite{DaSe:96}. In this paper we 
investigate
the power spectra (PS) of ``Barkhausen noise'' in the
zero temperature non-equilibrium RFIM and subsequent scaling relations near
the disorder induced critical point.   
Many of the methods will be applicable to
other systems with crackling noise \cite{SDM:01},
arising from the system's proximity to an underlying non-equilibrium
critical point. Within the RFIM the pulses of the Barkhausen signal are 
understood as collective events
(dubbed avalanches) in which many magnetic domains (spins) flip in time as the external field 
is slowly ramped up or down.  The properties of the Barkhausen signal and corresponding scaling relations
are then equivalent to the statistical properties of the avalanches which are obtained by virtue of the system's 
proximity to a critical point.

The association of an avalanche in the RFIM with a pulse in the 
Barkhausen signal presents some practical complications in many
experimental systems: background noise and finite field sweep rates 
lead to merging avalanches and ambiguous definitions of pulses.
It therefore seems best to use spectral tools \cite{Weis:96,PW:98} 
to analyze the pulse train as a whole, rather than, say, 
trying to extract pulse time, or pulse size distributions.
The Power Spectra (PS) that will be studied in this paper is defined as 
the absolute amplitude square of the Fourier Transform of the voltage 
time series $V(t)$. The voltage $V(t)$ obtained in Barkhausen noise 
measurements is the voltage induced in a pickup coil as a function 
of time $t$. The signal $V(t)$ is proportional to the
change in magnetization during the microscopic time interval
from $t$ to $t+\Delta t$, where $\Delta t$ is the time it takes a single
spin to flip.  An alternative and very useful approach to the study of 
Barkhausen noise at finite
sweep rate is provided by the ABBM model \cite{ABBM:90t}, which is similar
to the mean field approximation of the RFIM. As we will see, for
both the ABBM model and the RFIM the high frequency scaling behavior of
the PS is independent of the sweep rate and scales with a universal
exponent $1/(\sigma \nu z)$ having a value
of $2$ within the ABBM model and becomes around $1.77$ within the RFIM.

The organization of the paper is as follows. 
In section~\ref{SECT__Ising} we define the zero temperature, non-equilibrium
RFIM and review previous
results on the scaling behavior of the avalanche size distribution and
magnetic hysteresis curves near the underlying disorder induced
critical point for an adiabatically slowly increased external magnetic
field. In section~\ref{SECT__ADCase} we introduce the PS,
and give scaling predictions for the PS as a function
of the amount of disorder near the critical point, again 
for the adiabatic case.
In section~\ref{SECT_finite-sweeprate} we generalize the results 
on the PS to the case where the external field is swept
at a finite rate and give results for the expected critical exponents
obtained from Widom scaling collapses of the numerical simulation
results. In section~\ref{SECT__Inter} an application of the
PS to detect lack of causality in the order of spin flips during an 
avalanche is presented. 
Finally, in section~\ref{SECT__Concl} we summarize
the results obtained and discuss related questions to be addressed in the 
future.

\section{Hysteresis Modeled with Ising Spins}\label{SECT__Ising}

In the zero-temperature RFIM  
magnetic domains are represented by Ising spins
($s_{i}=\pm 1$) on a hyper-cubic $d$-dimensional lattice. Spins interact 
ferromagnetically with
their nearest neighbors with a strength $J$, and are subjected to a
homogeneous external field $H(t)$. Structural disorder is included by a 
quenched random 
field $h_i$ at each lattice site, with a Gaussian distribution
\be\label{Gauss_Dis}
\rho(h_{i})=\frac{1}{2\pi R} e^{-\frac{ h_{i}^2}{2R^2}} \ ,
\ee
where $R$ parametrizes the strength of the disorder. Spins interact with
an Ising Hamiltonian
\be\label{RFIM_Ham}
{\cal H}=-J\sum_{\langle i,j \rangle} s_{i} s_{j}-
\sum_{i}(H+h_{i})s_{i} 
\ee
The notation $\langle \rangle$ implies summation over nearest neighbor 
pairs only. 
The temperature is set to zero and each spin is aligned 
with its local effective field $h^{\mbox{eff}}_{i}$, defined by 
\be\label{LocField}
h^{\mbox{eff}}_{i} = J\sum_{j \in <i>} s_{j} +H +h_i \ ,
\ee
where $<i>$ denotes the nearest neighbors of site $i$.
The system is studied far from thermal equilibrium, {\it i.e.} it is typically
{\it not} in the ground state.

A dynamics is introduced into the system as follows: Initially the
external magnetic field is at $-\infty$, so that all spins are 
pointing down ($s_i = -1$ for all $i= 1,..., N $).
The magnetic field is then slowly increased.
The effective field $h_{i}^{\mbox{eff}}$ at each site is computed. 
If $h^{\mbox{eff}}_{i}$ at site $i$ changes sign,
the corresponding spin  $s_i$ at that site is flipped. 
A spin flip may trigger neighboring spins to flip as well, thus leading to an 
avalanche of spin flips, which is the analogue of a Barkhausen pulse in 
real magnets.

In this paper, the external field will be ramped up and down at a 
constant rate 
\be\label{def__rate}
\Omega \equiv \frac{dH}{dt} \ .
\ee

\section{The Adiabatic Limit}\label{SECT__ADCase}

\begin{figure}[htcb]
\centerline{
\epsfig{file=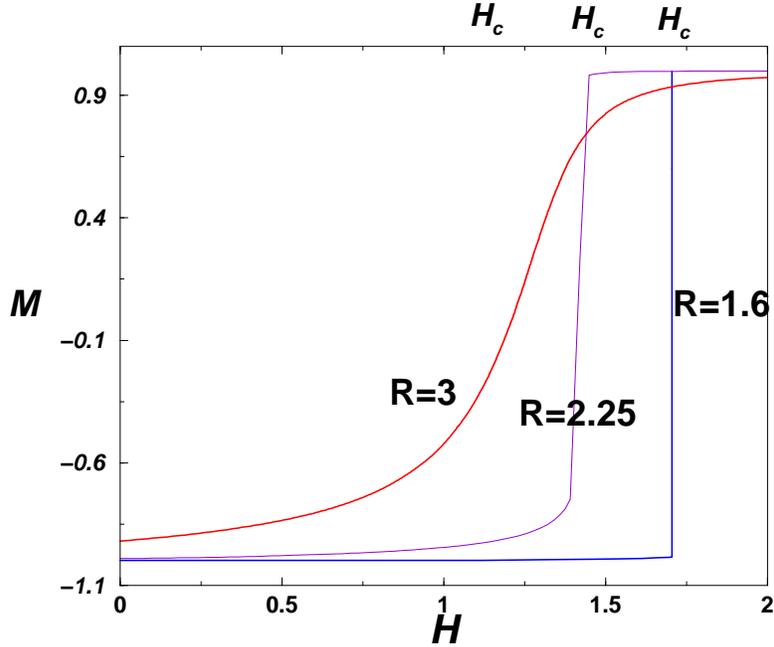,width=4in}}
\caption{Right branches of hysteresis loops (for increasing $H$) 
for high disorder $R=3$, close to critical disorder
$R=2.25$ and low disorder $R=1.6$. Results are for $L=200$ at
$\Omega=10^{-6}$.}
\label{fig__Trans}
\end{figure}

The zero sweep rate limit 
$\Omega \rightarrow 0$ is the adiabatic limit. In this limit
the external 
magnetic field is kept fixed during an avalanche. Only after the avalanche 
has come to a halt, the external field is increased until it triggers
the next avalanche. 
Combined analytical \cite{DaSe:96} and numerical approaches
\cite{KPDRS:99} 
have shown that in this case there is a {\it disorder driven} dynamical phase 
transition at a critical disorder $R_c$ ($R_c=2.16$ in three dimensions,
for the RFIM with nearest neighbor interactions and Gaussian disorder,
in units of the exchange coupling $J$), separating
a low disorder regime ($R<R_c$) characterized by hysteresis
loops with a macroscopic jump ($\Delta M$) in the magnetization $M$,
from
a high disorder regime ($R>R_c$), in which the hysteresis loops
look smooth, see Fig.~\ref{fig__Trans}. Here the magnetization 
is defined as $M \equiv (\sum_{i=1}^N s_i)/N$,
where $N=L^d$ is the total number of spins in the system, $L$ its linear
size and $d$ the dimension. The simulation results reported in this
paper are exclusively for $d=3$.

\begin{table}[h]
\centerline{
\begin{tabular}{|c|c|c|c|c||c|c|}\hline
\multicolumn{1}{|c|}{$\beta$} &
\multicolumn{1}{c|}{$1/\sigma$}& 
\multicolumn{1}{c|}{$\tau$} & 
\multicolumn{1}{c|}{$\beta\delta$} &
\multicolumn{1}{c||}{$\nu$} & \multicolumn{1}{c}{$H_c$}  
& \multicolumn{1}{c|}{$M_c$} 
\\\hline
 $0.035(30)$ & $4.20(30)$ & $1.60(6)$ & $1.8(2)$ & $1.4(2)$ & $1.435(4)$ & 
$0.9(1)$ \\\hline
\end{tabular}}
\caption{Critical exponents and critical fields defined in the text,
obtained from simulations of the adiabatic case in $d=3$.
The values of $H_c$ and $M_c$ are not universal. Values quoted 
correspond to a RFIM with n.n interactions and with a Gaussian distribution
of random fields. From \cite{PDS:97}.}\label{Tab__defexpo}
\end{table}

The jump $\Delta M $ in the 
magnetization for $R<R_c$ scales to zero as 
\be\label{jump_del}
\Delta M \sim (R_c-R)^\beta \ ,
\ee
where $\beta$ is a universal prediction of the model, table~\ref{Tab__defexpo}.
At the critical disorder ($R=R_c$) each branch of the saturation 
hysteresis loop has one point ($M_c(H_c)$) where the slope diverges, 
$dM/dH (H_c) \rightarrow \infty$: near that point the magnetization
is described by a power law of the form 
\be\label{Mag_del}
| M(H) - M_c | \sim | H-H_c |^{1/\delta}
\ee
where $\delta$ is another universal prediction for experiments 
(table~\ref{Tab__defexpo}).

The apparently smooth parts of the hysteresis loops really consist of many
microscopic steps not resolved in Fig.~\ref{fig__Trans}. These steps are 
the avalanches of spin flips, in analogy to Barkhausen pulses in
real materials. 
For $R > R_c$
the distribution of avalanche sizes $D(S,R,H)$ (which is proportional
to the probability to observe an avalanche of $S$ spin flips at disorder
$R$ and external field $H$ in response to a small increase in $H$) 
scales as 
\be
D_{int}(S,R,H) \sim  S^{-{\tau}} 
{\bar{\cal D}}_+(S^\sigma r, h/r^{\beta\delta}) \, ,
\ee
with $r \equiv (R-R_c)/R_c$, $h \equiv (H-H_c)/H_c$, and 
${\bar{\cal D}}_+$ is a {\it universal scaling function} \cite{PDS:99}.
There are similar scaling forms for the avalanche correlation function,
the cluster correlation function, and other quantities \cite{PDS:99}.
The correlation length $\xi$ scales as the diameter 
of the largest avalanche of the power law scaling regime of $D(S,R,H)$:
\be
\xi(r,h) \sim r^{-\nu} \Xi(h/r^{\beta\delta}) \ ,
\ee
where
$\Xi$ is a universal scaling function \cite{PDS:99}. Values for the 
exponents are given in table~\ref{Tab__defexpo}.
Interestingly, numerical simulations indicate that the 
``critical region'' is remarkably large: almost 3 decades of power
law scaling in the avalanche size distribution remain when 
measured at a disorder $R$ that is $40\%$ 
away from the critical point. At $2\%$ away one extrapolates seven decades
of scaling \cite{PDS:99}. This may explain why in many experiments
it does not even seem to be necessary to tune the disorder to see the
critical power law scaling over several decades: the samples used
may just fall into this large critical region.

\subsection{Power Spectra}\label{SECT_adiabatic}

Each avalanche, or Barkhausen pulse, has an internal structure reflecting
its evolution with time, see Fig.~\ref{fig__Time_Series}. We denote with
$V(t)$ either
the voltage measured in a Barkhausen noise experiment, or the number of spins 
triggered in our model
during the same microscopic time interval of 
length $\Delta\tau$, which
is the time it takes a single spin to flip. In our model we set $\Delta\tau=1$.
The PS is defined as the amplitude square of 
the Fourier Transform of the voltage
\be\label{Power_Spec}
P(\omega)=|\sum_{t=t_0}^{t_0+T} e^{i \omega t} V(t)|^2 .
\ee
There is some freedom in choosing the initial
time $t_0$ and the total duration of the transform $T$. We will
provide concrete criteria to fix these parameters.

In the strict adiabatic limit ($\Omega=0$), avalanches are separated
in time by an infinite time, and therefore, only the PS of an isolated
avalanche may be rigorously defined. The PS spectra may be defined from
the following limiting process: The frequency 
$\Omega_A$ is defined as the largest sweep-rate such that no 
two avalanches 
overlap in time. The frequency $\Omega_A$ will be finite in a finite system
and will approach zero with increasing volume. The adiabatic limit can
then be defined as the infinite volume result of the PS as computed with
the sweep rate $\Omega_A$. A general scaling form for 
the PS in the adiabatic case may be derived from generalizing the
result of Kuntz and Sethna \cite{KS:00} to the present situation
\be
\label{PSpectr}
P(\omega,h,r,L) \sim 
\omega^{-1/\sigma \nu z}  
{{\cal F}}(\omega^{-1/(\nu z)} r,  h/r^{\beta\delta}, L r^\nu)L^{d} 
|\Delta M | \ ,
\ee
where $\Delta M\equiv M(t_0+T)-M(t_0)$.
It was found in \cite{KS:00}, that at criticality, 
for $\tau < 2 $ (as in the RFIM at critical field and disorder), 
$P(\omega) \sim \omega^{-1/\sigma\nu z}$ for large $\omega$.
On the other hand, for $\tau > 2$, 
at the critical point, for large $\omega$, one finds 
$P(\omega) \sim \omega^{(\tau-3)/\sigma\nu z}$.
The scaling form ${\cal F}$ is therefore more conveniently split into 
two contributions \cite{KS:00} so that the general form for the PS becomes
\bea
\label{P-of-h}
P(\omega,h,r,L) &\sim& 
\omega^{-1/\sigma \nu z}
{{\cal F}}_1( \omega^{-1/(\nu z)}r,h/r^{\beta\delta}, L r^\nu) L^d|\Delta M|
\\\nonumber
&& + \omega^{(\tau-3)/\sigma \nu z} L^{\frac{\tau-2}{\sigma \nu z}} 
{{\cal F}}_2 (\omega^{-1/(\nu z)}r,h/r^{\beta\delta}, L r^\nu) 
L^d|\Delta M| \, 
\eea
where ${{\cal F}}_1$ and ${{\cal F}}_2$ are universal scaling functions, which
become finite for large frequencies.

\section{Finite Field Sweep rate}
\label{SECT_finite-sweeprate}

The above scaling forms were all obtained for zero field sweep rate.
In the following section we take into account that in experiments
typically the external magnetic field
is ramped up and down with a constant finite sweep 
rate $\Omega \equiv dH/dt >0 $.
\begin{figure}[htcb]
\centerline {\epsfig{file=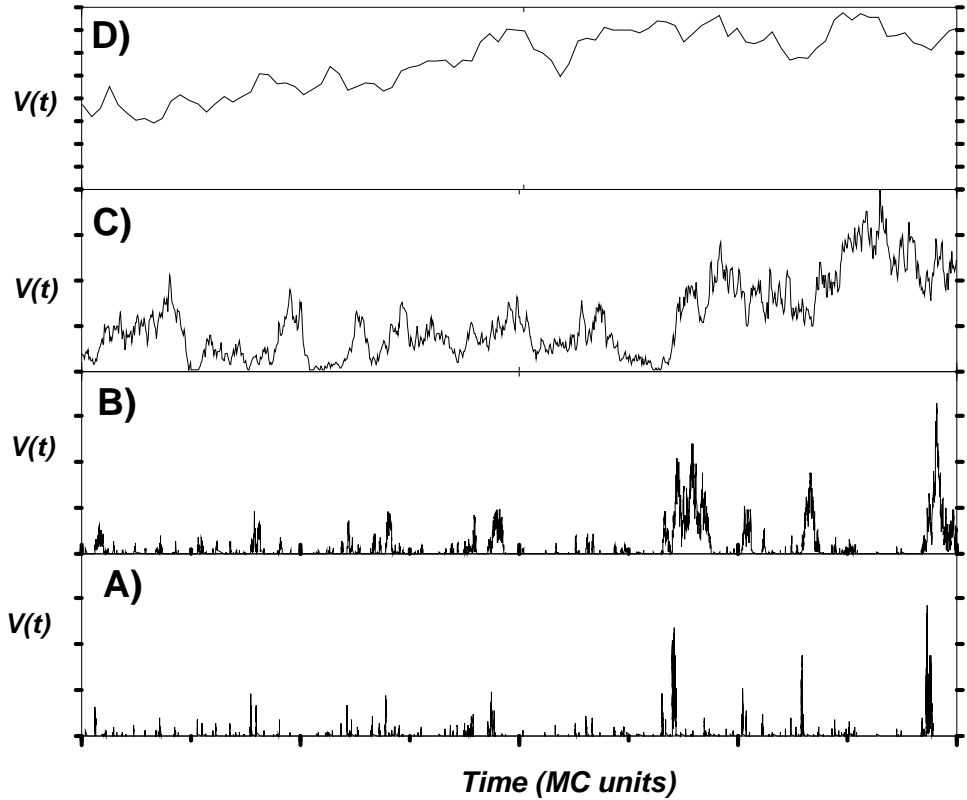,width=5in}}
\caption{Time series for the noise signal in a ferromagnetic
system. This corresponds to a time series for sweep
rates of increasing magnitude from A to D.}
\label{fig__Time_Series}
\end{figure}
In order to compare sweep rates at different volume sizes, it
is convenient to define
\be\label{FF_H_indV}
\Omega = \frac{v}{N} ,
\ee
where $N$ is the number of spins in the actual system and $v$ defines a 
size independent sweep rate.
In contrast to the adiabatic case ( $\Omega \rightarrow 0$ )
where the external magnetic field is kept fixed during an avalanche, at 
finite field sweep rate it can trigger new avalanches before currently 
running ones have petered out, see Fig.~\ref{fig__Time_Series}.

Interestingly, the Barkhausen noise time series at relatively fast
sweep rates $\Omega$ (as in case $D$ in 
Fig.~\ref{fig__Time_Series}, where basically no separate avalanches 
can be identified since most individual avalanches overlap in time)
has the same PS as the low driving sweep rate cases $A$, $B$ 
or $C$, at least for high frequencies $\omega$, and therefore results
in the same high frequency noise scaling exponents. This highlights
the usefulness of the power spectra: it allows to characterize Barkhausen 
noise in situations where individual avalanches cannot be observed separately.

To anticipate the effect of finite sweep rate we note the general form of the
power spectra of an ensemble of independent pulses given in \cite{Mazz:62}
where it is assumed that there are no size correlations between pulses:
\be\label{indep-pulses-powerspectrum}
P_{total}(\omega )=a(H)\Omega \left[ \left\langle P(\omega )
\right\rangle +2\left| \left\langle \Phi (\omega )\right\rangle 
\right| ^{2}Re\left( \frac{\int ^{\infty }_{0}D(\Delta T)
e^{i\omega \Delta T}d\Delta T}{1-\int ^{\infty }_{0}
D(\Delta T)e^{i\omega \Delta T}d\Delta T}\right) \right] \ ,
\ee
where \( \left\langle \right\rangle  \) is the average over all the individual
pulses, \( \Phi (\omega ) \) is the Fourier transform of an individual 
pulse and \( D(\Delta T) \) is the distribution of time intervals between
independent, successive nucleation events ({\it i.e.} spin flips that
in the adiabatic limit would be triggered only by an increase in the
external magnetic field rather than some neighboring spin flips). 
$a(H)$ is the number of nucleation events per unit field
increase, yielding $a(H) \Omega \mathcal{T}$ nucleations in a time 
span $\mathcal{T}$. 

\begin{figure}[tcb]
\centerline {\epsfig{file=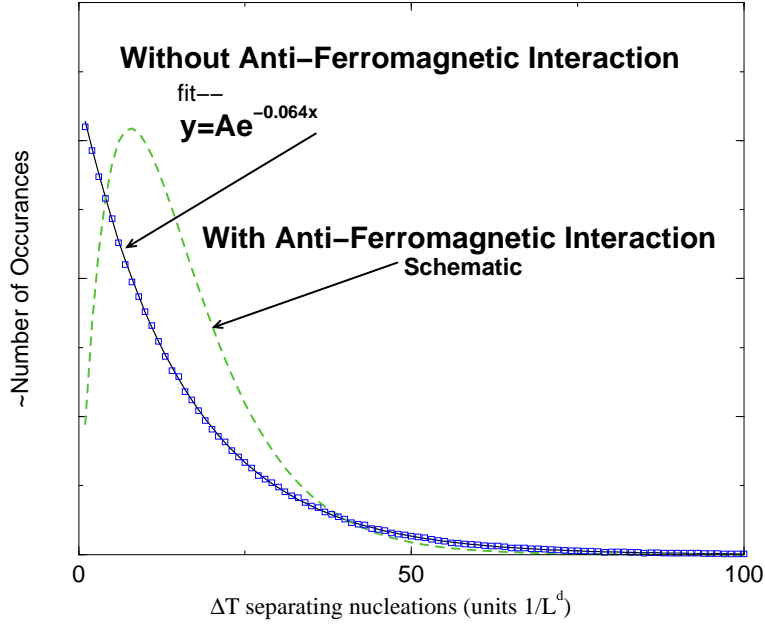,width=4in}}
\caption{Plot of the distribution of time intervals $\Delta T$ separating
successive nucleation events. The continuous
line is a fit to a Poissonian distribution. For comparison, a schematic
figure for an anti-ferromagnetic interaction is shown.}
\label{fig__Pois}
\end{figure}

As shown in Fig.~\ref{fig__Pois}, for the RFIM with only n.n. 
interactions \( D(\Delta T) \) is Poissonian, 
leaving the second term in brackets zero. 
Hence the resulting PS is simply proportional to the sum of
the individual PS. This holds as long as any simultaneously
propagating collective events
do not overlap in {\it space} (otherwise the pulses can no longer be considered
independent \cite{DW:02}). Conveniently, this implies that the PS can
be used to determine many scaling
properties of the simultaneously propagating adiabatic avalanches at 
high sweep rates. 

It is important to note here that the introduction of long range 
(anti-ferromagnetic) dipole-dipole 
interactions, as present in soft magnetic materials,
would lead to distinctly non-Poissonian
\( D(\Delta T) \) which results in a nontrivial sweep rate dependence 
in \( P_{tot}(\omega ) \). We leave a detailed study of this dependence 
for the future. The results of this paper are expected to apply to
hard magnets and other systems, in which long range dipolar
fields are negligible due to sample geometry and other factors.

\subsubsection{The determination of $H_c$}\label{SubSECT__Above_HC}

Before testing the above scaling relations Eq.~\ref{P-of-h}, 
%(\ref{P-integrated}), 
and Eq.~\ref{indep-pulses-powerspectrum}, we 
first need a criteria to pick up a sensible initial time $t_0$ and
total length $T$ of the Fourier transformed Barkhausen train
to compute the PS (see Eq.~\ref{Power_Spec}). 
In order to apply Eq.~\ref{P-of-h} at the critical field ($h=0$) we
choose the Fourier transformed interval such that
\be\label{Pow_Ab_cond3}
H_c \in [H(t_0),H(t_0+T)] \ .
\ee
Furthermore, in order to have sensible statistics, one would like
to include as many spins flips as possible in the time series,
while keeping the deviation from $H_c$ small enough so that 
Eq.~\ref{P-of-h} with $h=0$ is still applicable. We typically choose the
intervals such that $h\equiv (H-H_c)/H_c < 0.05$. This way,
it is found that a reasonable fraction of the spins are included in the
time series and the range of recorded events spans all sizes.

\subsubsection{How fast is fast?}
\label{SubSECT__Above_fast}

\begin{figure}[tcb]
\centerline {\epsfig{file=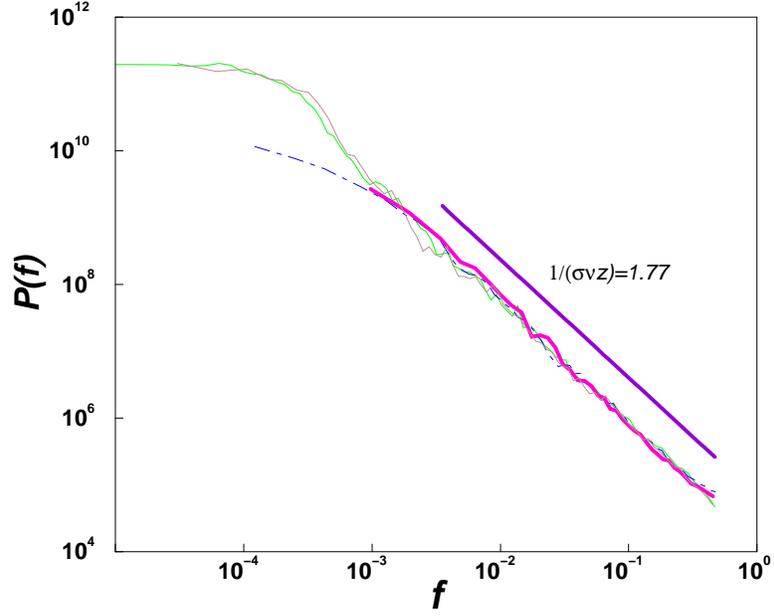,width=4in}}
\caption{Plot of the PS (for $f\equiv\frac{\omega}{2\pi}$)
as a function of $\Omega$ at $R_c$ for
$L=100$. Result includes $\Omega=10^{-9}$ and 
$\Omega=10^{-6}$ (thin solid lines), 
$\Omega=10^{-4}$ 
(dot-dashed line) and 
$\Omega=10^{-3}$ (wide solid line). The fit $P(f)\sim f^{-1/(\sigma \nu z)}$
corresponds to $\Omega=10^{-9}$ and has been separated for proper 
visualization.} \label{fig__Pow_FF}
\end{figure}

We now compute the critical sweep-rate $\Omega_c$, above which one 
observes changes of the PS due to spatially overlapping 
simultaneous events.
Since the nucleations of avalanches are random
in time ({\it i.e.} Poisson distributed) Eq.~\ref{indep-pulses-powerspectrum} 
implies that the 
adiabatic scaling result for $P(\omega)$ will hold as long as the spin
flip order is maintained for the adiabatic avalanches.  That is, as long 
as we are superimposing the adiabatic avalanches to construct the time 
series we are assured of obtaining a power spectra
that is the superposition of the PS of the individual pulses. 
The flip order begins to change when, during the propagation of an avalanche,
new avalanches are triggered within the same volume that the initial 
avalanche would have eventually covered.
The increased sweep-rate leads to ``parallel processing'' and
the avalanche that would have taken $T$ time steps to complete now takes 
$T'$ such that $T' < T \mbox{ for all} \mbox{  } \Omega> \Omega_c(T)$.
The field sweep rate at which  the expected number of spins 
flipped in avalanches triggered within the volume
of the primary avalanche is some finite fraction of the primary avalanche 
size is given for an avalanche of duration T as \cite{DW:02} 
\be\label{seeded_inside}
\Omega_c(T)= \frac{\xi^{d_{f}(\tau - 2)}}{T}
\ee
Since the power spectra is a superposition of all the individual avalanches there will not be any change
in the PS as long as
\be\label{seeded_inside_crit}
\Omega < \Omega_c \equiv \frac{\xi^{d_{f}(\tau - 2)}}{T^{*}}
\ee
where $T^{*}$ is the largest duration observed which scales as $T^{*} \sim 
\xi^{z}$.  Since $\xi \sim L$ 
if $r^{-\nu}>L$ or 
$\xi \sim r^{-\nu}$ if $r^{-\nu}<L$ the critical field sweep rate 
has two scaling forms
\be\label{crit_sweep_L}
\Omega_{c} \sim  L^{d_{f}(\tau - 2 - \sigma \nu z)} \mbox{  for  } \xi >> L
\ee
\be\label{crit_sweep_xi}
\Omega_{c} \sim \xi^{d_{f}(\tau - 2 - \sigma \nu z)} \mbox{  for  } \xi << L.
\ee
We note here that if $\tau - 2 - \sigma \nu z > -1$, as is the 
case in the RFIM, then
even if the avalanches are compact, {\it i.e.} $d_{f}=d$, the onset 
of sweep rate effects happens at a field sweep
rate much faster than the largest sweep rate at which one can
still measure separate pulses
(since $\Omega$ is measured in units of $1/L^d$).  
Fig.~\ref{fig__Pow_FF} confirms the applicability of the  adiabatic result to
a large range of external sweep rates $\Omega< \Omega_c$.

Furthermore even for $\Omega > \Omega_c$ 
we can derive a frequency $\omega_a(\Omega) \equiv \Omega\xi^{d_{f}(2-\tau)}$ 
so that for $\omega > \omega_a(\Omega)$ the power spectra 
are again described by the adiabatic results \cite{DW:02}.
Below this frequency the PS follows a cutoff function as a 
result of the sweep rate decreasing the duration of the larger
avalanches (see Fig.~\ref{fig__Pow_FF}).
For $\omega << \Omega $ one finds
the asymptotic $\omega \rightarrow 0$ result 
\be\label{Pow_freq}
\lim_{\omega \rightarrow 0} P(\omega )= L^{2 d} |M(t_0+T)-M(T)|^2
\equiv L^{2d} |\Delta M|^2  \ .
\ee
Recalling the scaling relation \cite{DaSe:96}
\be\label{scaling_reg}
\Delta M \sim r^{\beta} {\cal M}(h r^{-\beta \delta},L r^{\nu}) \ ,
\ee
near the critical point, in the adiabatic case ($\Omega\rightarrow0$ first)
\be\label{Pow_freq_new}
\lim_{\omega \rightarrow 0} P(\omega )= 
L^{2(d-\beta/\nu)} \mbox{  for r $\rightarrow 0$ ,  $h\rightarrow 0$} .
\ee
Let us recall that while in \cite{KS:00},
the limit of small $\omega$ is taken after
the adiabatic $\Omega \rightarrow 0$ limit has been performed, in our 
simulations we will be working at
finite sweep rates and  therefore, our PS will always satisfy 
Eq.~\ref{Pow_freq} correctly (with non-zero $\Delta M$). It
will be shown, however, that in $d=3$ dimensions, $\beta/\nu$ 
will be very close to zero and therefore, even the adiabatic limit 
scales very approximately as $L^6$.

\subsubsection{Critical exponents}

The PS is now obtained from numerical simulations
and the scaling prediction Eq.~\ref{P-of-h} 
in section~\ref{SECT_adiabatic} is used to extract critical exponents.
The results presented involve averages over several, usually about 16,
disorder realizations.

\begin{figure}[hp]
\centerline {\epsfig{file=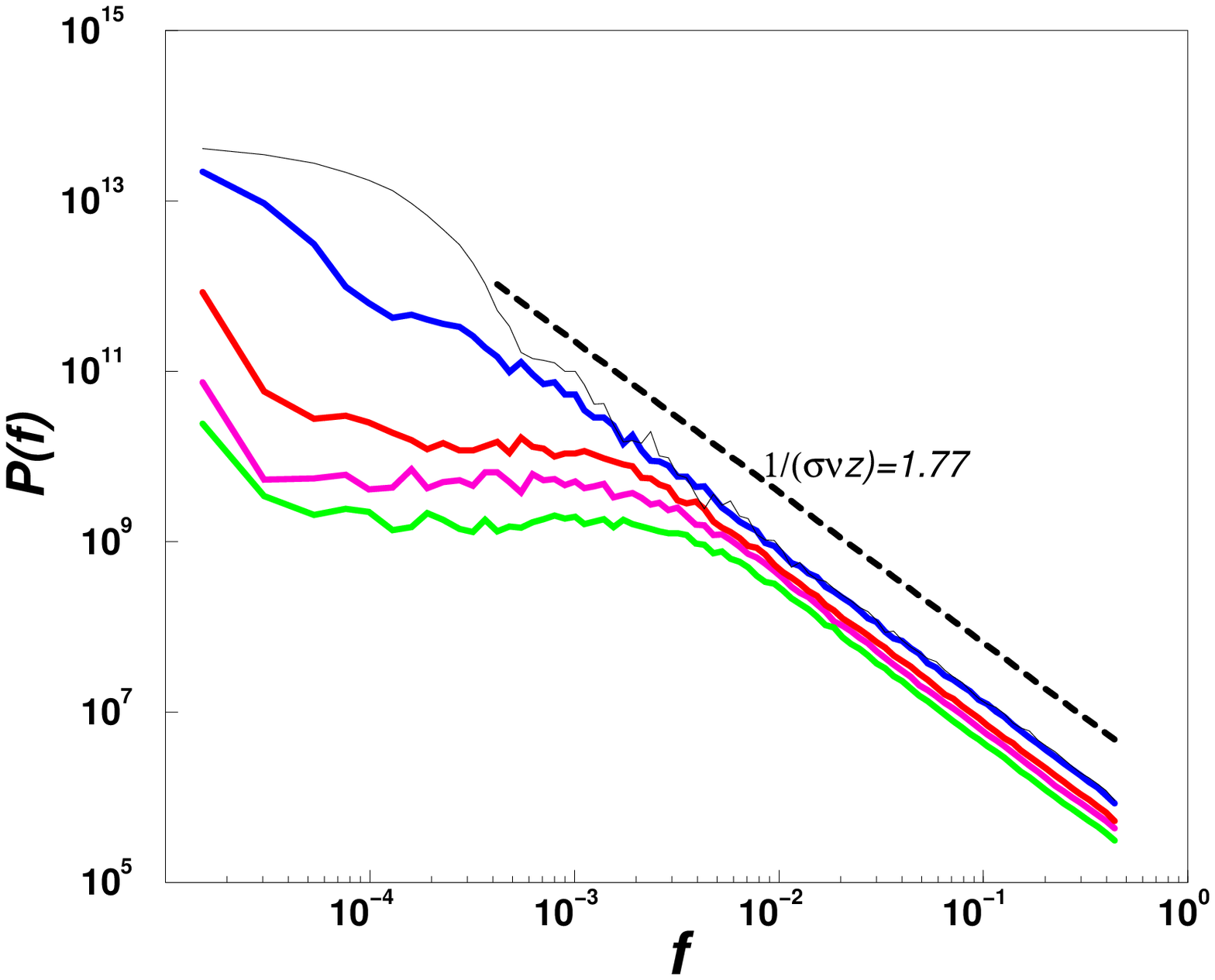,width=3.1 in}}
\caption{Power Spectra for ($f=\frac{\omega}{2 \pi}$) 
at different disorder values
$R=2.30$, $R=2.50$ $R=2.60$, $R=2.70$. The thin line corresponds 
to $R=2.20 < R_c$, and the
dashed line is the fit to $R=2.30$ (it has been shifted for proper 
visualization). Results correspond to $L=200$,
 $\Omega=10^{-6}$, and $H$ in a small interval around $H_c$ with
$(H-H_c)/H_c < 0.05$.}
\label{fig__Pow_Above_RC}

\centerline{\epsfig{file=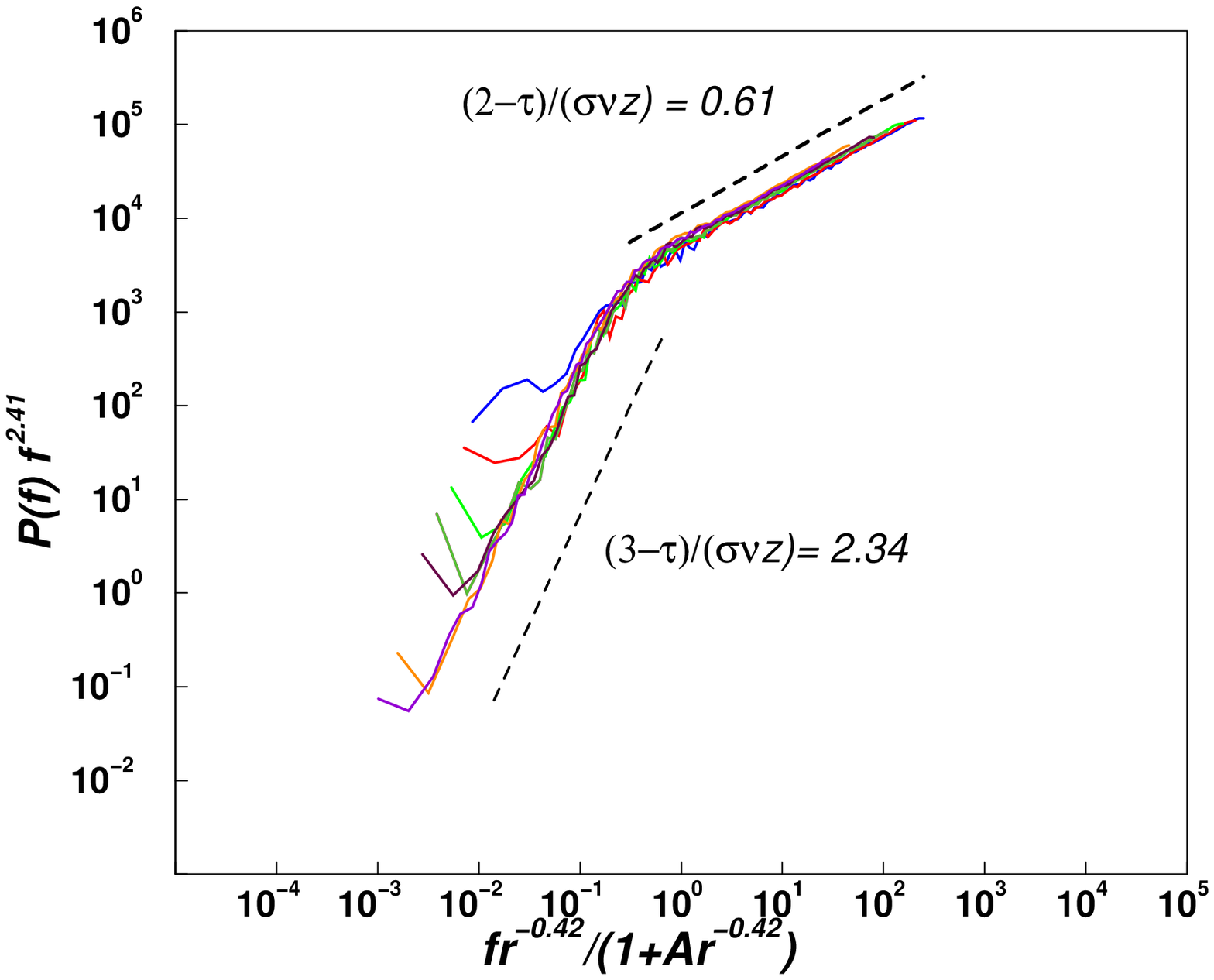,width=3.1 in}}
\caption{Power Spectra (for $f=\frac{\omega}{2\pi}$) collapse for
 disorder values $R=2.3$ to $R=2.7$ 
at $L=200$, $\Omega=10^{-6}$, according 
to Eq.~\ref{Pow_Ab_Sca2}. The value of $A$ is a non-universal 
parameter dependent on the size of the system L. ($A \sim L^z$).}
\label{fig__Pow_Above_FR}
\end{figure}

Besides the external driving frequencies, there are two additional natural 
frequencies in the system defining the characteristic times of the largest 
events in the system:
\be\label{omega_r}
\omega_r \equiv \frac{1}{\xi(r)^{z}} \ ,
\ee
where $\xi(r)$ is the correlation length, and 
\be\label{omega_L}
\omega_L \equiv \frac{1}{L^{z}} \ ,
\ee
where $L$ is the finite linear size of the system.

For future convenience, we introduce a simple continuous 
function $\omega_c$ that 
asymptotically approaches the values of  $\omega_r$ and $\omega_L$, as 
$\xi/L \rightarrow 0$ and $\xi/L \rightarrow \infty$, respectively:
\be
\label{Crit_Freq}
\omega_c\equiv\frac{1+\xi(r)^{z}/L^z}{\xi(r)^{z}}= \omega_r+\omega_L \ .
\ee
Other definitions with the same assymptotics can also be used. 

In Fig.~\ref{fig__Pow_Above_RC} the PS for $H$ from a small interval around 
$H_c$ is shown as a function of the 
disorder strength for fixed system size. As the disorder approaches $R_c$, 
the system size $L$ becomes the characteristic length scale of the system 
implying that $\omega_L$ takes over $\omega_r$ as the characteristic
cut-off frequency, and one observes a very large region of power law
scaling (about four decades for $L=200$). For larger disorder $R> R_c$,
it becomes apparent that $\omega_c \sim \omega_r \sim r^{\nu z}$  
in agreement with the behavior suggested in Eq.~\ref{P-of-h}.

At large frequencies $P(\omega) \sim \omega^{-1/\sigma \nu z}$,
as expected, with 
\be\label{Exp_Above_1}
1/(\sigma \nu z)=1.77(4) \ ,
\ee
where the error bar merely indicates the dispersion around the mean value
from the different fits to Eq.~\ref{P-of-h} corresponding 
to the different disorder parameters and system sizes considered.

More information can be obtained from a scaling collapse of the PS 
for different disorders. 
Setting $h=0$ in the adiabatic scaling form Eq.~\ref{P-of-h} we can 
make an Ansatz for the specific form of the scaling function ${\cal F}$
to arrive at the scaling form (assuming $\tau < 2$, as discussed) for
$\frac{\xi}{L}<1$:
\be\label{Pow_Ab_Sca2}
P(\omega)=(\frac{\omega}{\omega_L})^{-\frac{3-\tau}{\sigma \nu z}}{\cal Q}(
\frac{\omega}{\omega_c}) |\Delta M| L^{d+1/{\sigma \nu}} \ .
\ee

The factor of $|\Delta M|L^d $ is a normalization factor resulting
from our specific definition of $P(\omega)$. Its particular form in
the adiabatic case has been
given in Eq.~\ref{Pow_freq_new}.
The asymptotic behavior of the function ${\cal Q}$ is given by simple
power laws (with $x=\frac{\omega}{\omega_c}$):
\be\label{Pow_Above_ScaQ}
{\cal Q}(x) = \left\{ \begin{array}{c}
  x^{\frac{2-\tau}{\sigma \nu z}} \ \ \ \ , \ \, x >> 1 \\ 
  x^{\frac{3-\tau}{\sigma \nu z}} \ \ \ \ , \ \, x << 1  \end{array} \right.
,
\ee
where the scaling for $x>>1$ implies the expected behavior 
$\omega^{-1/(\sigma \nu z)}$ at large frequencies. Plugging
Eq.~\ref{Crit_Freq} and Eq.~\ref{Pow_Above_ScaQ} into Eq.~\ref{Pow_Ab_Sca2},
for $x <<1$, one obtains:
\be\label{Pow_Above_Cond4}
P(\omega)\sim \left(\frac{\xi}{L}(1+(\frac{\xi}{L})^z)\right)^
{\frac{3-\tau}{\sigma \nu z}} \ , \ \omega << \omega_c \ ,
\ee
{\it i.e.} the PS becomes independent of the frequency. This is
a consequence of the fact that the system is above criticality, there is a 
minimum cut-off frequency $\omega_r$, which corresponds to the finite 
duration of the largest avalanche in the system, with no characteristic
events having larger durations. This result is reflected
by the low frequency plateau at large 
disorder in Fig.~\ref{fig__Pow_Above_RC}.

\begin{figure}[hp]

\centerline{\epsfig{file=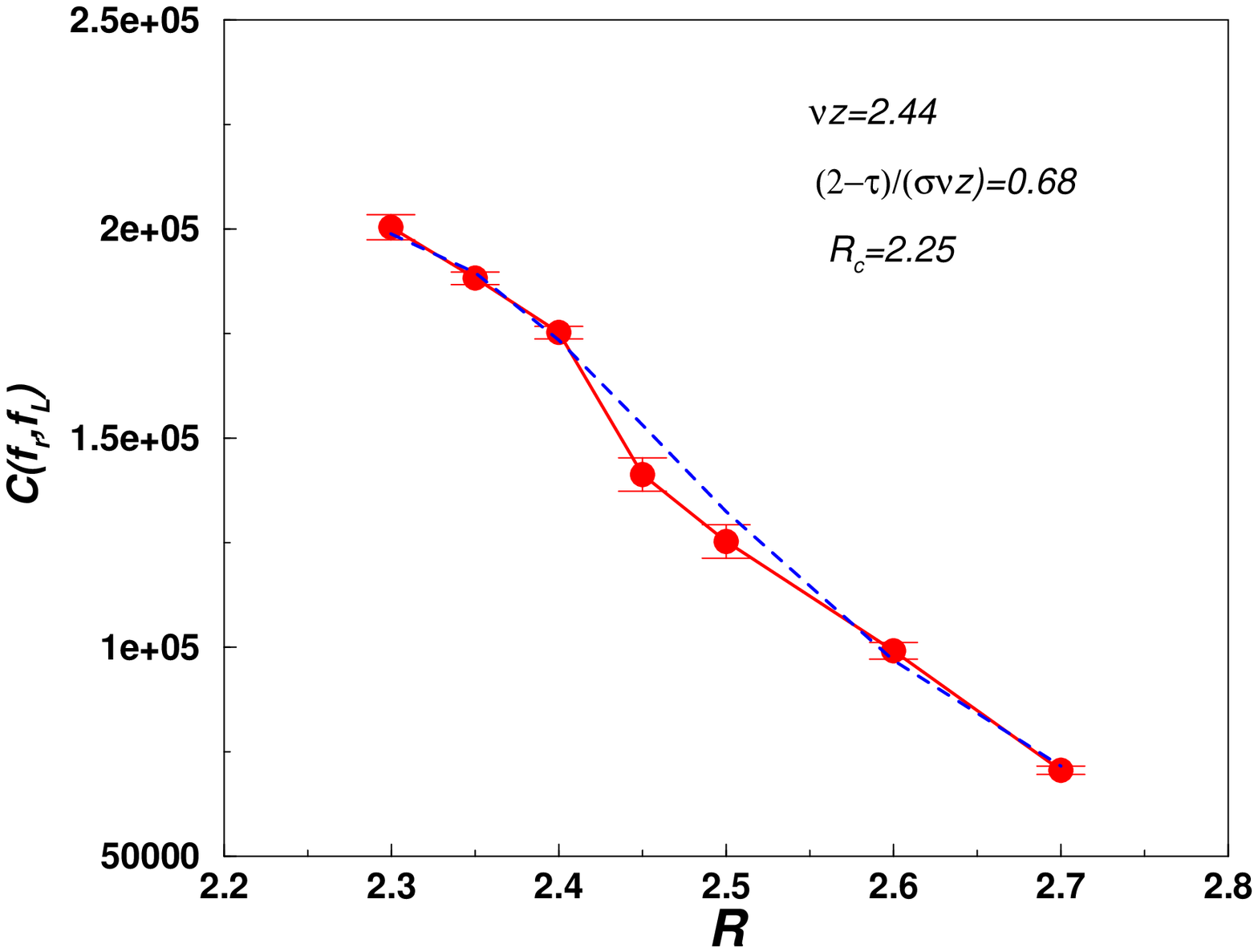,width=3.8 in}}
\caption{The coefficient 
$C(f_r,f_L)\equiv \frac{C(\omega_r,\omega_L)}{(2\pi)^{1/\sigma\nu z}}$, see
Eq.~\ref{Pow_Above_Sca3} as a function of
disorder $R$ for $L=200$ and $\Omega=10^{-6}$ (grey dots). The dashed line 
corresponds to a fit according to Eq.~\ref{Pow_Above_Sca3}.}
\label{fig__Pow_Above_C}

\centerline{\epsfig{file=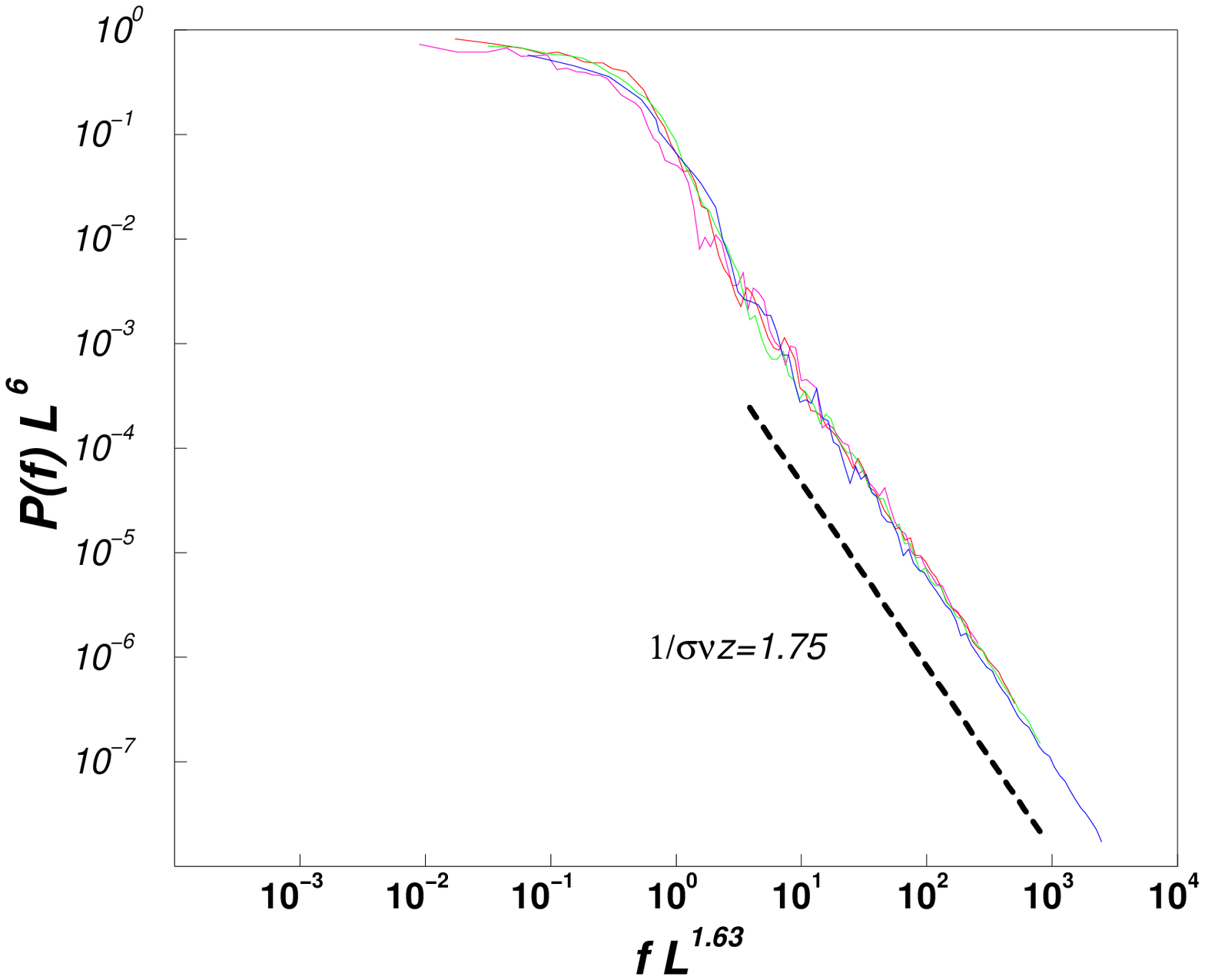,width=3.5 in}}
\caption{Finite size scaling collapse of Power Spectra 
(for $f=\frac{\omega}{2\pi}$) at $R=R_c$ 
for system sizes $L=50$ to $L=200$, (L=50,75,100,150,200),
collapsed according to Eq.~\ref{Pow_Ab_Sca1}.}
\label{fig__Pow_Above_FL}

\end{figure}

The collapse obtained using the scaling form Eq.~\ref{Pow_Ab_Sca2}
is plotted in 
Fig.~\ref{fig__Pow_Above_FR}. For the particular system size
 shown ($L=200$), the collapse extends up to about five decades and 
it only fails 
for very low frequencies that are of order or smaller than the (small)
nonzero external sweep rate $\Omega$ in the simulations.

From the collapse one obtains the critical exponents
\be\label{Exp_Above_3}
\frac{1}{\nu z}=2.38(10) \ , \ \frac{3-\tau}{\sigma \nu z}=2.41(10) \ , \
\frac{2-\tau}{\sigma \nu z}=0.61(9) \ , \
\ee
with error bars estimated as previously.

It is found that $L^d \Delta M$ is numerically very close to $L^d$, and
from this and the scaling form Eq.~\ref{scaling_reg} one derives that $\beta$ 
is close to zero within the accuracy of our calculation. 

Additional information may be obtained by further exploiting the large 
frequency behavior in Eq.~\ref{Pow_Ab_Sca2}, $\omega >> \omega_c$. We can
write 
\bea\label{Pow_Above_Sca3}
P(\omega)&=&C(\omega_r,\omega_L) \omega^{-1/(\sigma \nu z)} \ \ \ \ \omega >> \omega_c \nonumber\\ 
&&C(\omega_r,\omega_L)=
\bar{A}(L)\left(\frac{r^{-\nu z}}{1+(\frac{\xi}{L})^z}\right)^{\frac{2-\tau}
{\sigma\nu z}} \ ,
\eea
One can fit the coefficient $C$ for different disorder values 
(see Fig.~\ref{fig__Pow_Above_C}), where $L$ and therefore
$\bar{A}(L)$ is a constant).
The fit is excellent at the
two ends of the curve and becomes slightly inaccurate on the region
$\xi/L \sim 1$, which is a consequence of the approximation made in 
selecting the interpolation
function for $\omega_c$ given in Eq.~\ref{Crit_Freq}. Since the exponents 
are computed from the two ends, one obtains
\be\label{Exp_Above_4}
\frac{1}{\nu z}=2.44(10)  \ , \ \frac{2-\tau}{\sigma \nu z}=0.68(8) \ , \
\ee
%and also
%\be\label{RC_Above_2}
%R_c=2.25(3) \ ,
%\ee
which provides a cross check of the validity of the results obtained in
the collapse.
It is also interesting to compute the correlation length
\be\label{xi_Above_1}
\xi(R)=9.68\left(\frac{R-R_c}{R_c}\right)^{-1.40} \ .
\ee
The large exponent $\nu \simeq 1.4(1)$ is partly responsible for the large
scaling region of the non-equilibrium RFIM \cite{DaSe:96}.

\subsubsection{Finite Size Scaling}

The analysis of the PS as a function of the linear system size 
$L$ at criticality $r=h=0$
is also very interesting. From Eq.~\ref{P-of-h} one derives
the adiabatic scaling form
\be\label{Pow_Ab_Sca1}
P(\omega)L^{1/(\sigma\nu)+d} |\Delta M|=
P(\omega)L^{d-\beta/\nu+1/(\sigma \nu)}=
{\cal G}(\omega L^z) \ ,
\ee
with the universal scaling function ${\cal G}(x) \sim 1$ for 
$x \rightarrow 0$ and 
${\cal G}(x) \sim x^{-1/(\sigma \nu z)}$ for $x \rightarrow \infty $. 
Fig.~\ref{fig__Pow_Above_FL} shows that the data
collapse very well to the assumed relation for all the
included frequencies (about five decades of scaling for $L=200$). 
The resulting exponents are
\be\label{Exp_Above_2}
z=1.63(10) \ \ , \ \  1/(\sigma \nu z)=1.75(6) \ \ , \ \ 
\theta/2+1/\sigma\nu=3.0(5)
\ee
where the exponent relation $\theta=d-\beta/\nu-1/(\sigma \nu)$
\cite{DaSe:96} has been used and 
the error bars indicate the dispersion from 
the values obtained as a consequence of disorder fluctuations.

\subsection{Below $R_c$}\label{SubSECT__Below}

\subsubsection{General Forms}\label{SubSECT__Below_Gen}

The hysteresis loops of the RFIM below the critical disorder $R_c$ 
have a macroscopic jump in the magnetization at a critical field
$H_c(R)$ where many spins flip almost
simultaneously in an event that extends over a finite
fraction of the system.
The jump grows with decreasing disorder.
Studying the critical behavior of the RFIM below $R_c$ gets complicated
by the fact that even if the correlation length becomes smaller than
the system size, there is always an event which is sensitive to
the system size, and 
in previous studies, large finite size effects have made the analysis at
$R < R_c$ difficult.
The PS below but near $R_c$ for $\omega > \Omega$ can be collapsed using 
the scaling form
\be\label{Pow_Below_sca1}
P(\omega)=\phi(\frac{\omega/\omega_L}{1+\omega_r/\omega}) |\Delta M|
L^{1/(\sigma \nu)+d} .
\ee
with 
$\phi(x) \sim x^{-1/(\sigma\nu z)}$ for $x >> 1$. The latter result states
that short-time propagation of the avalanches proceeds in the same way, 
whether above or below (but close to) the critical disorder. At zero disorder,
the PS is a constant:
\be\label{PS_zz}
\lim_{R \rightarrow 0} P(\omega)=L^{2d} \ .
\ee
According to Eq.~\ref{Pow_freq}, this also scales as the limit of the PS for 
$R>0$ for $\omega << \Omega$. The total duration $T_{\Delta M}(L,R)$ of 
the jump
in the magnetization for $R<R_c$ decreases for decreasing $R$. This leads to
an increasing cut-off frequency 
\be\label{ctt_off}
\omega(r,L) \sim \frac{1}{T_{\Delta M}(L,R)} \ ,
\ee
with $P(\omega) \sim 1/\omega^{1/(\sigma \nu z)}$ for $\omega > \omega(r,L)$ 
and $P(\omega) \rightarrow C L^{d+1/(\sigma\nu)}|\Delta M|$ 
with $C$ a constant, for 
$\Omega << \omega << \omega(r,L)$ and $P(\omega)\rightarrow L^{2d}$ for
$\omega << \Omega$.

Note that the scaling form Eq.~\ref{Pow_Below_sca1} is really a 
special case of the general form
Eq.~\ref{P-of-h} for $\tau <2$ derived above $R_c$ with different 
scaling functions ${\cal F}_1$ and ${\cal F}_2$.

Far below $R_c$ for very large systems we expect Eq.~\ref{Pow_Below_sca1} 
to be replaced by a 
scaling form that involves the critical exponents of single interface
depinning \cite{BKD:98,JRob:92}, some of which are very close to
the exponents of the critical point studied in this paper (for
example, $\frac{1}{\sigma\nu z}$ is the same within error bars) \cite{KS:00}. 

For the numerical analysis we note that
the general considerations discussed in subsections~\ref{SubSECT__Above_HC}
and \ref{SubSECT__Above_fast} concerning choosing the interval so that it 
includes $H_c(R)$, its width and the range of sweep rates, applies in the
regime below $R_c$ as well.

\subsubsection{Critical Exponents}\label{SubSECT__Below_Crit}

\begin{figure}[hp]
\centerline {\epsfig{file=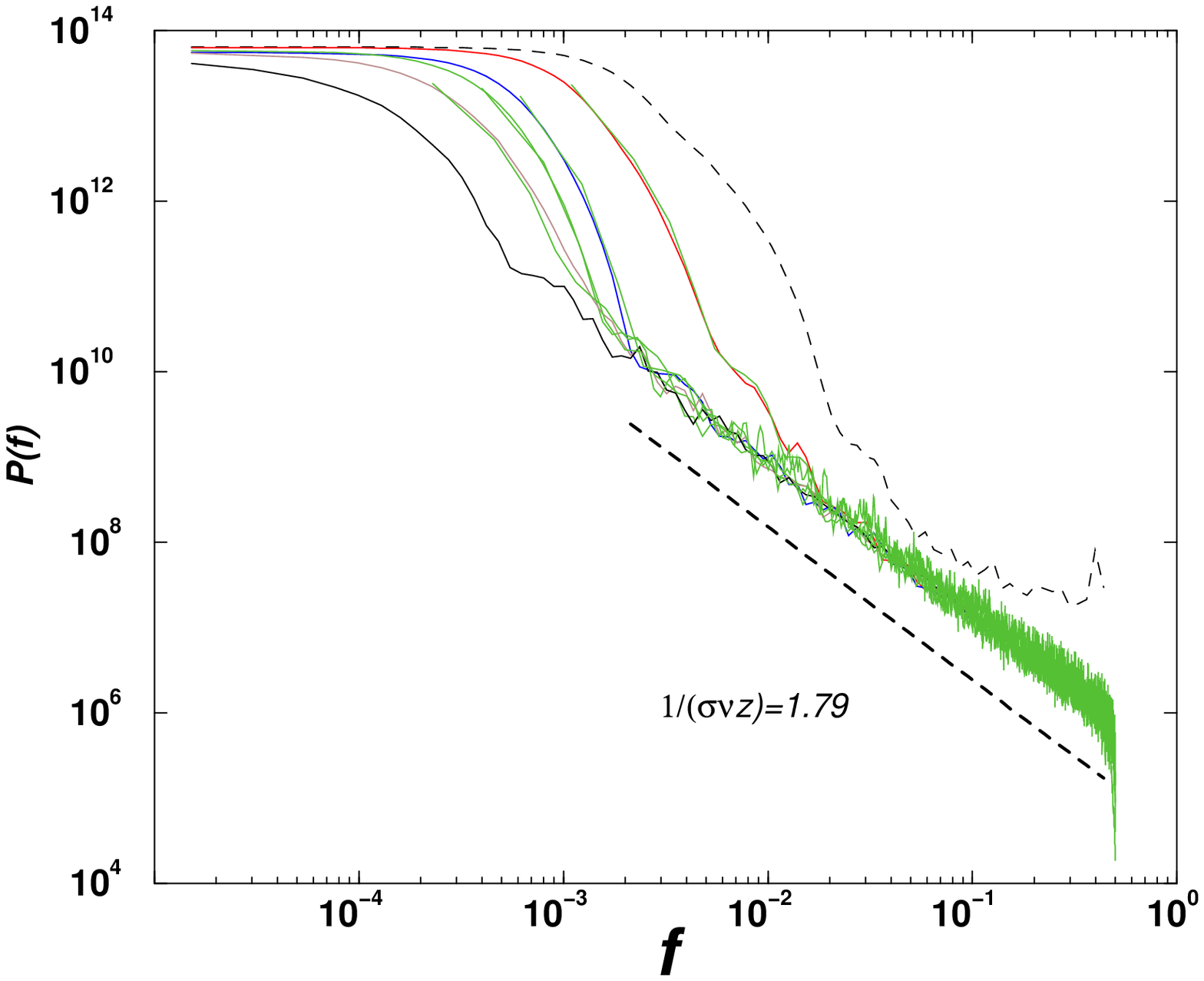,width=3.4in}}
\caption{The Power spectra (for $f=\frac{\omega}{2\pi}$) for data 
below $R_c$ for $L=200$ and disorders between $R=2.2$ and $R=1.6$. $R=1.0$ 
is plotted with a thin dotted line, for $\Omega=10^{-6}$. All other
curves are plotted for
$\Omega=10^{-6}$ and $\Omega=0$ (adiabatic limit).
The fit (straight dashed line) is slightly shifted for proper 
visualization.}
\label{fig__Pow_Below_fr}

\centerline {\epsfig{file=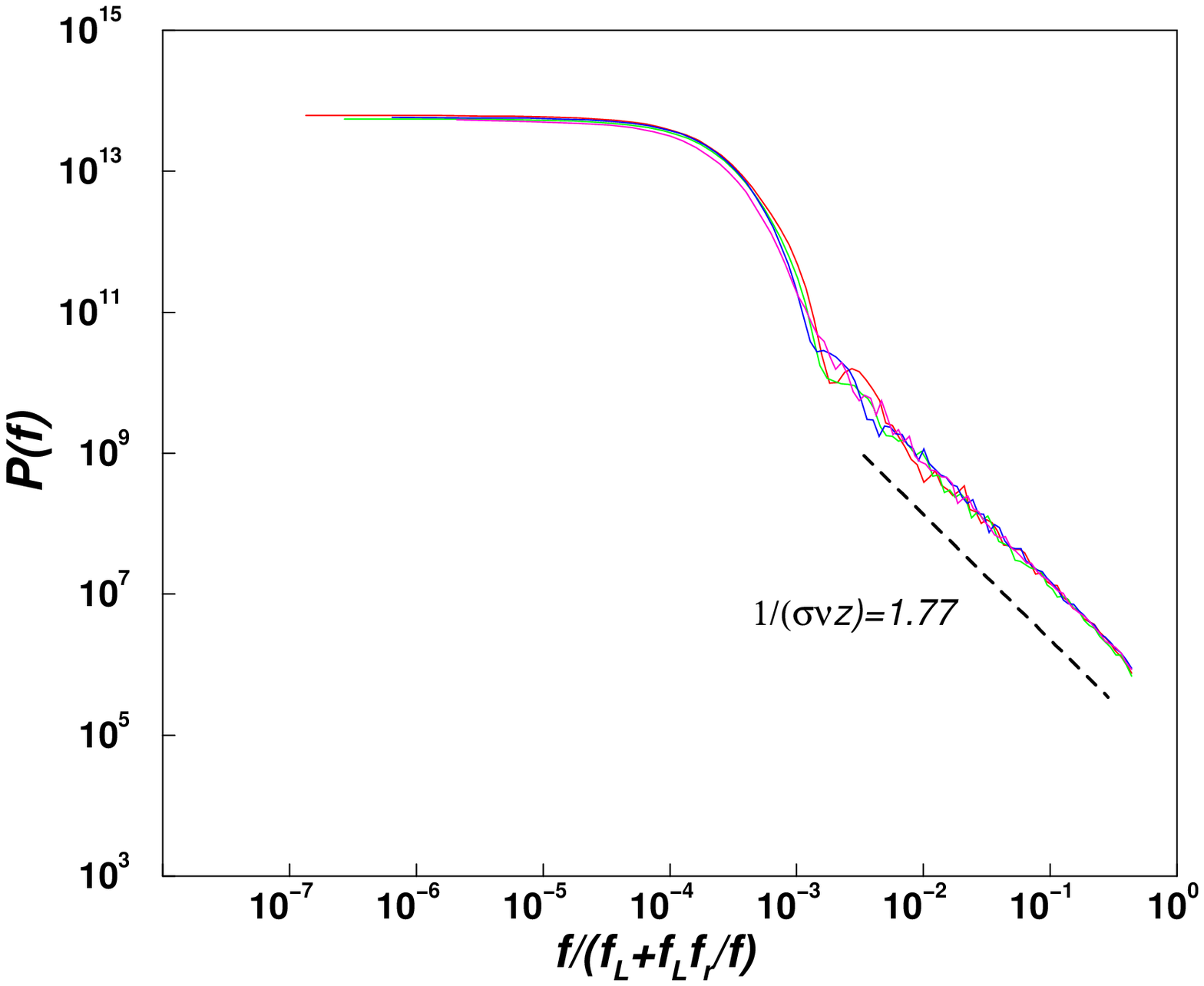,width=3.4in}}
\caption{Collapse of the Power spectra (for $f=\frac{\omega}{2\pi}$)
for disorders $R=2.2$ to $R=1.6$, for
$L=200$ and $\Omega=10^{-6}$. The data is collapsed 
according to Eq.~\ref{Pow_Below_sca1}, and $f_L\equiv \omega_L/2\pi$ and
$f_r\equiv \omega_r/2\pi$.}
\label{fig__Pow_Below_col}
\end{figure}

The first issue we investigate is the large frequency limit, 
which is described by Eq.~\ref{Pow_Below_sca1}.
In Fig.~\ref{fig__Pow_Below_fr}
the results of the PS are plotted for different disorder parameters $R<R_c$,
for a low sweep rate $\Omega=10^{-6}$ and in the adiabatic case ($\Omega=0$)
the fit to the large frequency behavior gives
\be\label{Pow_Below_exp1}
1/(\sigma \nu z)=1.79(5)  \ ,
\ee
with error bars defined as previously. 

Fig.~\ref{fig__Pow_Below_fr} shows very clearly,
particularly for the PS at $R=1 << R_c$, that 
for decreasing disorder $P(\omega)$ tends to a constant over larger and
larger regimes $\omega < \omega(r,L)$,
and consequently the large frequency region where the behavior 
Eq.~\ref{Pow_Below_sca1} holds, shrinks. 

Fig. \ref{fig__Pow_Below_col} shows a collapse of the PS for different
disorders according to Eq.(\ref{Pow_Below_sca1}) over about six decades 
of scaling for $L=200$, with the critical exponents
\be\label{Pow_Below_exp2}
\nu z = 2.38(10) \ , \  1/(\sigma \nu z)=1.77(5) \ .
\ee
In contrast to the analysis for $R$ above $R_c$,
the analysis below $R_c$ does not involve the
exponent $\tau$, related to the distribution of avalanche sizes.

More information below $R_c$ may be obtained from
the region $\xi(r) >> L$ where finite size effects dominate. Actually,
the PS may be collapsed exactly as in Eq.~\ref{Pow_Ab_Sca1}, and 
the exponents 
\be\label{Exp_Below_2}
z=1.63(10) \ \ , \ \  1/(\sigma \nu z)=1.75(6) \ ,
\ee
follow. Since the PS in this region becomes insensitive to the
correlation length $\xi$, the results are identical to
the ones above $R_c$.

\section{The Interpretation of the Power Spectra}\label{SECT__Inter}

The previous analysis has shown that the PS is sensitive
to the causality relations among spins. 
This point can be made more explicit by considering a different
dynamics than the causal dynamics implemented in this paper.  Let us consider
a standard metropolis dynamics, as if we were introducing temperature
into the model. The algorithm is as follows: Spins are
randomly selected\footnote{since the disorder is itself random, 
a sequential update has very similar effect as a random one, with
very similar PS}, 
and the standard Metropolis (or Glauber) acceptance/rejection check is
used to determine whether the spin is to be flipped. Let us 
recall that since we are considering the temperature to be zero, the
proposed move is accepted if the local energy $E_i\equiv -h_i^{eff}\sigma_i$ 
of the spin being
tested is positive and rejected if it is negative. After $N$ such
attempts to flip randomly chosen spins are performed, the time is increased 
by one unit. 
 
\begin{figure}[hp]
\centerline{\epsfig{file=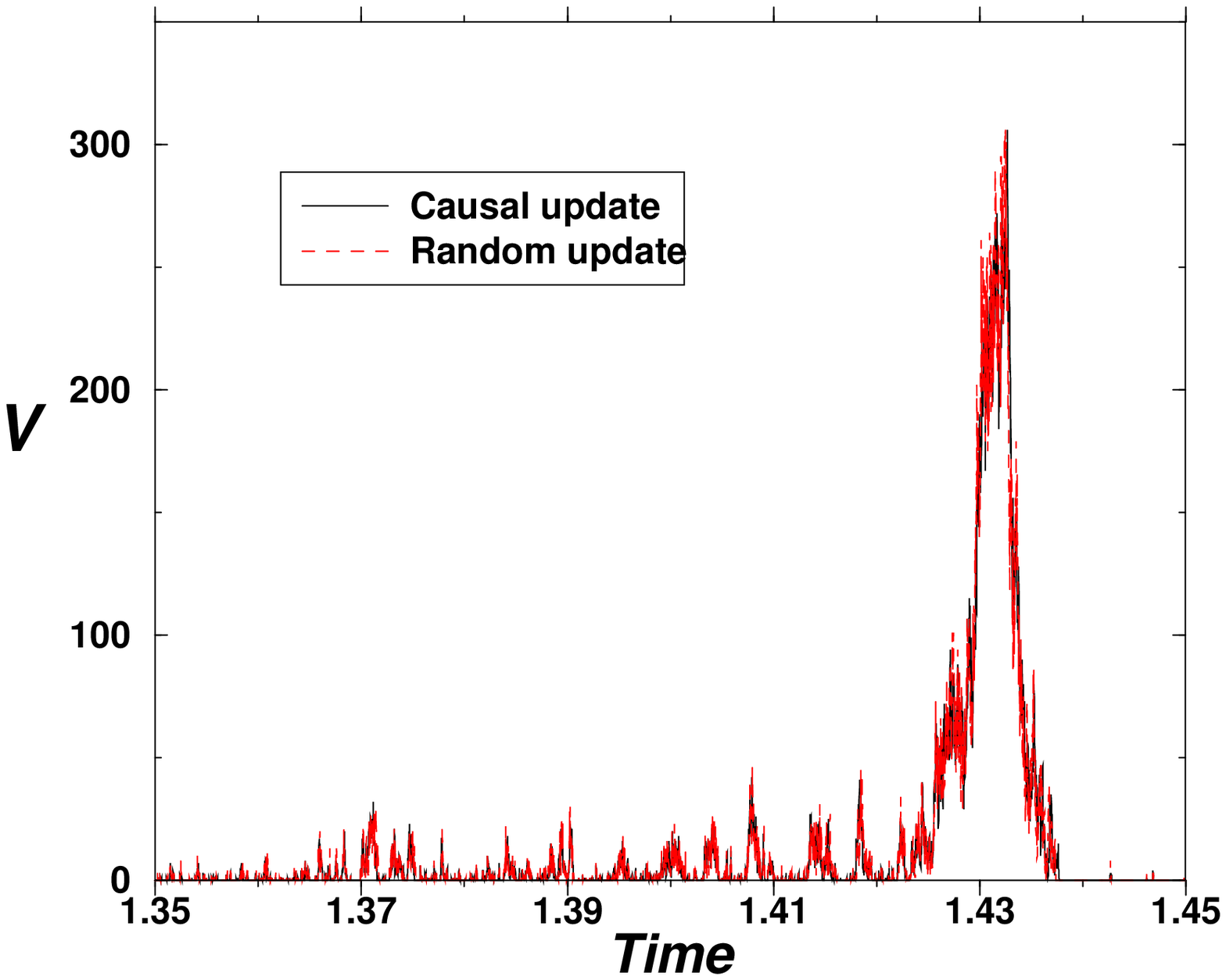,width=3.8 in}}
\caption{Voltage time series for causal and random updates for a single 
disorder realization $R=2.3)$.}
\label{fig__Pow__Caus_v_Random}

\centerline{\epsfig{file=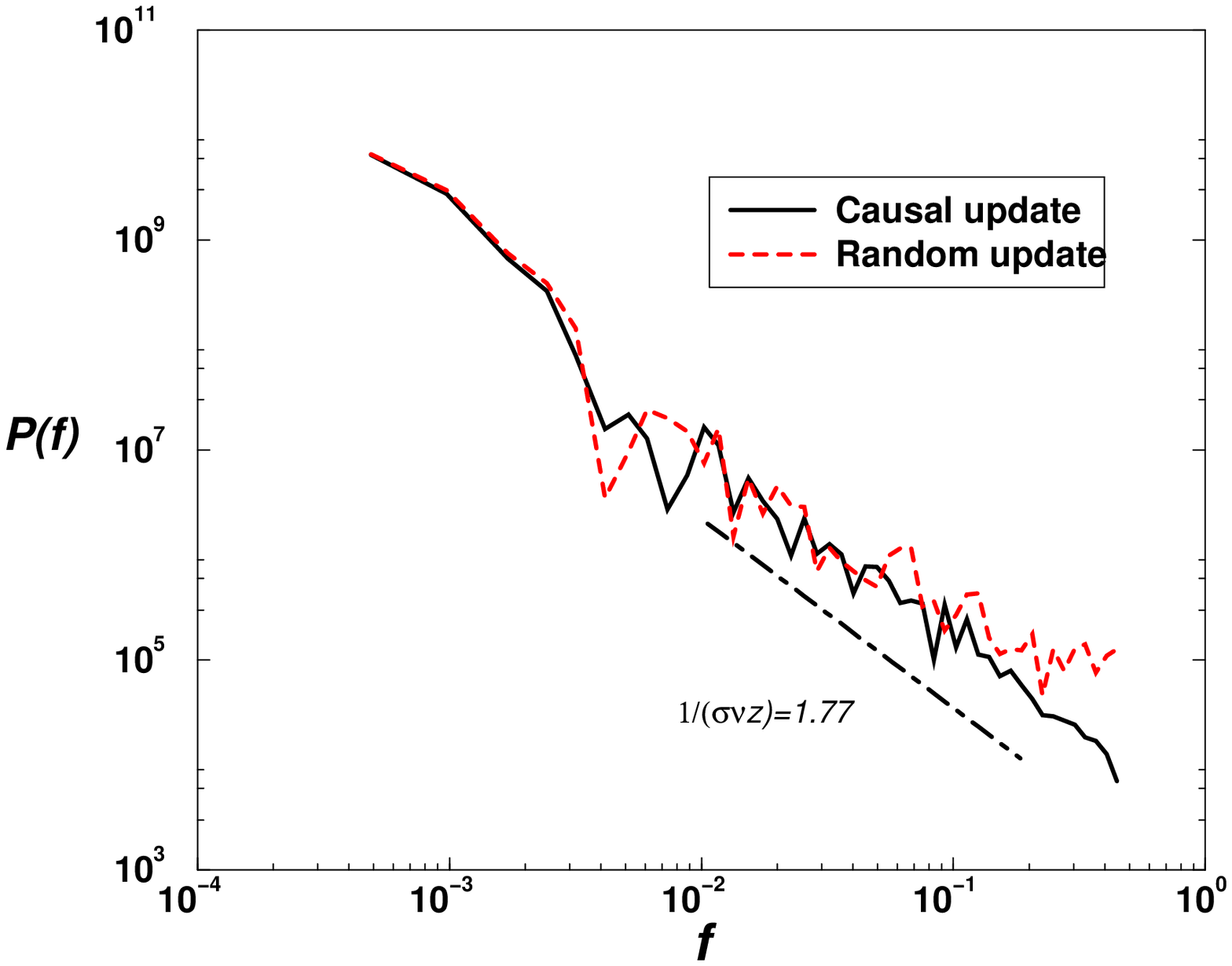,width=3.8 in}}
\caption{Power Spectra (for $f=\frac{\omega}{2\pi}$) of the previous
time series.}
\label{fig__Caus_v_Random}
\end{figure}

A spin can only flip when it is selected, and therefore, if that particular
spin is not selected by the random update, there will be a delay between the 
time when it is energetically favorable (which is the time when the spin 
would be flipped in the causal dynamics) and the time in which 
the spin actually flips. Spins flipping at earlier times as compared 
to the causal dynamics may also take place, since the random process may 
flip a spin seeding an avalanche, and within the same unit of time during the
$N$ attempts before updating the time by one unit,
some of the neighbors to this seeding avalanche spin, which become now 
energetically unstable, may be flipped. Within the causal dynamics those
spins would flip at later times.

In Fig.~\ref{fig__Pow__Caus_v_Random} the time series generated from these
two dynamics are shown and are extremely close. 
The PS, however, as shown in Fig.~\ref{fig__Caus_v_Random},
has a different high frequency behavior, and it is clearly sensitive to
the different dynamics.

Of course, time averages of the magnetization, etc.. will
be insensitive to the different dynamics, as is clear from the
time series Fig.~\ref{fig__Pow__Caus_v_Random}. Observables that
show some universality at large frequencies, as is the case
for the PS, become sensitive to the different dynamics of spins, even if
averaged over several disorder configurations.

The previous remarks are of interest for extensions of 
the previous study to more general situations, in particular
for considering temperature fluctuations when the system is 
out of equilibrium. Temperature would be naturally 
introduced into the system by considering Glauber dynamics with random
updates, as done in other simulations with Ising spins \cite{KWNR:00}.
Of course any algorithm to study temperature 
effects should meet the necessary requirements:
\begin{itemize}
\item Converge to equilibrium for very slow sweep rates at finite temperature.
\item Reduce to the causal dynamics for small temperatures at finite sweep 
rate.
\end{itemize}
It follows from our results that
the second condition is clearly violated for random updates (or any other
update uncorrelated with the dynamics of the system, like a sequential update).
The update must be chosen correlated with the dynamics, thus reflecting the 
causal nature of the events (the avalanches).  

\section{Conclusions}\label{SECT__Concl}

\subsection{Summary of results}

In this paper we have seen that the PS is an extremely
valuable observable for the study of  collective noise in physical
situations where the avalanche picture would no longer be applicable.

Besides its obvious theoretical interest, the PS becomes a 
powerful practical tool to compute critical quantities. 
We have computed the following exponents summarized in 
table~\ref{Tab__expo}.

\begin{table}[h]
\centerline{
\begin{tabular}{|c||c|c||c|c|}\hline
\multicolumn{1}{|c}{} &
\multicolumn{2}{|c|}{Above $R_c$}& 
\multicolumn{2}{c|}{Below $R_c$} \\\hline 
\multicolumn{1}{c}{Quantity} &
\multicolumn{1}{c}{PS} & \multicolumn{1}{c}{RFIM}  
& \multicolumn{1}{c}{PS} & \multicolumn{1}{c}{RFIM}
\\\hline
 $1/(\sigma \nu z)$ & $1.77(3)$   & $1.75(9)$  & $1.79(6)$ & \\\hline
 $z$ & $1.63(10)$  & $1.7(2)^{\ast}$ &     $1.63(10)$ & \\\hline
 $\nu z$  & $2.38(10)$  & $2.39(40)^{\ast}$ &2.38(10)  &   \\\hline
 $\frac{3-\tau}{\sigma \nu z}$ & $2.41(10)$ &$2.45(8)^{\ast}$ & & \\\hline
 $\frac{2-\tau}{\sigma \nu z}$ & $0.61(9)$  &$0.70(6)^{\ast}$ & & \\\hline
\end{tabular}}
\caption{Exponents computed and the ones known}\label{Tab__expo}
\end{table}

The exponent predictions lie well within the error bars of most
Barkhausen noise exponents from experiments quoted in the literature,
\cite{SDM:01}. Besides, we are able to obtain accurate exponents for the
region below $R_c$.
 
The exponent
quotes have been deliberately computed at relatively fast sweep rates, see
section~\ref{SECT_finite-sweeprate} for a detailed discussion, thus providing
conclusive evidence of the irrelevance of low enough finite field sweep rate 
for the PS. Finally, it has been shown that the PS is sensitive to the 
particular dynamics of the spins.

\subsection{Experiments and Outlook}

\begin{figure}[t]
\centerline {\epsfig{file=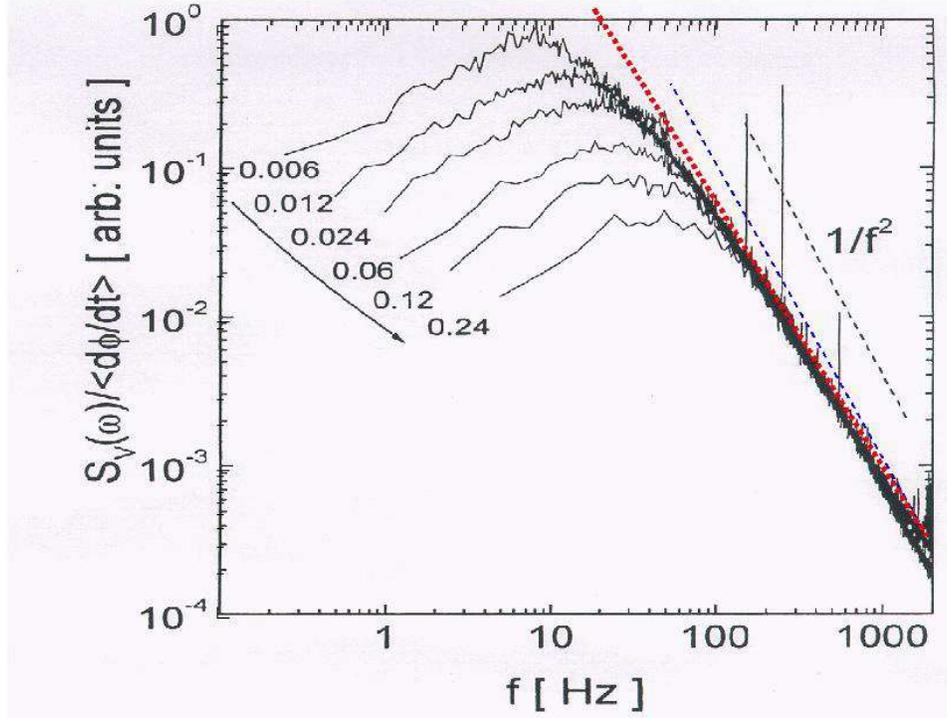,width=5in}}
\caption{Experimental results from \cite{ABBM:90e}. The dotted line is 
our prediction $1.78$. the dashed line is the result of $2$ from
the ABBM model \cite{ABBM:90t}.}
\label{fig__exp_ABBM}
\end{figure}

In Fig.~\ref{fig__exp_ABBM} our results are compared to the experiments
from \cite{ABBM:90e}, corresponding to a $FeSi$ sample. 
Our result for the large frequency dependence of 
the PS looks more plausible than the exponent $\frac{1}{\sigma \nu z}=2$, 
suggested
by the ABBM model \cite{ABBM:90t}. In \cite{QBa:01}, the exponent  
$\frac{1}{\sigma \nu z}=2$ is calculated reporting values in the range 
$\frac{1}{\sigma \nu z}=1.5-2$, in agreement with our results, but with
a too large uncertainty for a precise comparison.
Other experimental determinations in 
amorphous ribbons of $Fe_{64}Co_{21}B_{15}$ also seem to favor a 
value close for $\frac{1}{\sigma \nu z}$ close to $1.78$ 
\cite{DuZap:01}. For this particular data 
we expect that the frequency 
dependant bump in the PS is due to two things:
(1) the presence of long range anti-ferromagnetic (LRAF) interactions which 
(change the nucleation 
distribution from Poisson to something more complicated, see section
\ref{SECT_finite-sweeprate}, and) 
affect the low frequency of the power spectra;
(2) the fact that the larger avalanches take less time due to ``parallel processing'' discussed above also diminishes the adiabatic PS in the low frequency regime. While the full details of these effects were
not addressed in our current study we are presently investigating effects of LRAF interaction in
the RFIM.

In experiments on soft magnetic materials,  
Barkhausen noise seems to be due to the propagation of a 
single domain wall, and nucleation of new domains are rare events 
\cite{Zap:97}. Far below $R_c$, our model also exhibits domain wall 
propagation during the macroscopic jump $\Delta M$ in the hysteresis
loop. For small disorder $R$, the number of small nucleating events are
expected to be negligible, and in this case the PS for the jump at high
frequencies should match those of single interface propagation in the absence
of LRAF \cite{JRob:92,KS:00} (In some experimental systems the 
LRAF can be suppressed, for example by applying stress to the sample
\cite{DuZap:01}).
It is known that the critical exponents for single interface propagation
belong to a different universality class, although the
critical exponents characterizing the large frequency limit of the
PS turn out to be numerically very close \cite{SDM:01,Zap:97}. Therefore,
a very detailed study is needed to discriminate between these
two regimes (low and near critical disorder) in which the two models apply
to the PS.

The main problem for a detailed comparison of the predictions 
for the disorder induced critical point presented in this paper
with the experiments, 
is the lack of control over the structural disorder in 
experimental systems. Recent experiments on Co/CoO \cite{BIJPB:00} 
have shown that in these systems it is possible to control 
the amount of structural 
disorder experimentally. If Barkhausen noise is measured in these
disorder controlled samples, one could use the scaling forms derived
in this paper and get a very detailed understanding of the 
universal aspects of hysteresis, including detailed critical 
exponents near the critical disorder.

Other experimental systems for which such analysis could be extended
include acoustic emission in martensitic materials in response to
stress or temperature ramping. In fact, the Barcelona group 
\cite{VIV1:94,VIV2:95,VIV3:95,VIV4:98} has shown that crackling noise in
martensites displays critical scaling in avalanche size distribution and
the PS similar to the ones described in this paper for Barkhausen noise, 
although with different exponents and therefore belonging to a different
universality class.

The analysis presented in this paper certainly allows several important
applications which we have briefly discussed. First of all, the 
irrelevance of the sweep rate has very important practical consequences
for future calculations since by using a faster sweep rate 
one can compute the same quantities with less effort. This is
an important benefit in situations, as for example the inclusion
of temperature, where slow sweep rates are computationally very
demanding.

Another important aspect is how long range
dipole-dipole interactions may affect the
results, a subject briefly discussed in Fig.~\ref{fig__Pois}. We hope
to report more on that in the near future.

\bigskip
\bigskip
\bigskip

\noindent {\bf Acknowledgments} 
 
We thank P. Bellon, J.P. Sethna and M.B. Weissman for very 
claryfing discussions. We thank IBM for a very generous equipment award
that made extensive numerical simulations for this work possible.
The work by K.A.D, A.T. and R.A.W has been supported by the Materials
Computation Center, grant NSF-DMR 99-76550, and NSF grant
DMR 00-72783. K.A.D. also gratefully acknowledges support from an A.P.
Sloan fellowship.

\medskip 
 
\newpage  

\bibliography{RFIM} 

\end{document}